\documentclass{iaus}
\usepackage{graphicx}

\def\xmm{{\sl XMM-Newton}}

\def\chandra{{\sl Chandra}}
\def\Chandra{{\sl Chandra}}

\def\gtrsim{\lower 2pt \hbox{$\, \buildrel {\scriptstyle >}\over
{\scriptstyle \sim}\,$}}
\def\lesssim{\lower 2pt \hbox{$\, \buildrel {\scriptstyle <}\over
{\scriptstyle \sim}\,$}}

\def\ovii{O~{\scriptsize VII}}

\title[~~X-ray sources in galaxies] 
{Sources of X-rays from galaxies}

\author[Q. Daniel Wang]   
{Q. Daniel Wang$^1$}

\affiliation{$^1$Astronomy Department, University of Massachusetts, USA \\ email: {\tt wqd@astro.umass.edu}} 

\pubyear{2011} 
\volume{284}  
\pagerange{}
\jname{The Spectral Energy Distribution of Galaxies}
\editors{R.J. Tuffs \&  C.C.Popescu, eds.}
\begin{document}

\maketitle

\begin{abstract}
Galactic X-ray emission is a manifestation of various high-energy
phenomena and processes. The brightest X-ray sources are typically
accretion-powered objects: active galactic nuclei and low- or high-mass X-ray binaries. Such objects with X-ray luminosities of $\gtrsim 10^{37} {\rm~ergs~s^{-1}}$ can now be detected individually in nearby galaxies. The contributions from fainter discrete sources (including cataclysmic variables, active binaries, young stellar objects, and supernova
remnants) 
are well correlated with the star formation rate or 
stellar mass of galaxies. The study of discrete X-ray sources is
essential to our understanding of stellar evolution, dynamics, and
end-products as well as accretion physics. With the subtraction of the
discrete source contributions, one can further map out truly diffuse X-ray emission, which can be used 
to trace the feedback from active galactic nuclei, as well as from stars, both young and old, in the form of stellar winds and supernovae. The X-ray emission efficiency, however, is only about 1\% of the energy input rate of the stellar feedback alone. The bulk of the feedback energy is most likely gone with outflows into large-scale galactic halos. Much is yet to be investigated to comprehend the role of such outflows in regulating the ecosystem, hence the evolution of galaxies. Even the mechanism of the diffuse X-ray emission remains quite uncertain. A substantial fraction of the
emission cannot arise directly from optically-thin thermal plasma, as commonly
assumed, and most likely originates in its charge exchange 
with neutral gas. These uncertainties underscore our poor understanding of
the feedback and its interplay with the galaxy evolution. 


\end{abstract}

\firstsection 
\section{Introduction}
Why are X-rays from galaxies interesting? X-ray emission from a galaxy typically arises from high-energy (high temperature or relativistic) processes in extreme conditions. 
Such processes are important to understand in their own right, but are
also believed to play a major role in shaping the structure and evolution of galaxies via various
types of feedback (e.g., Tang et al. 2009;	
	Oppenheimer et al. 2010). In particular, the X-ray emission
        effectively traces diffuse hot plasma, which may dominate the
        volume and account for the bulk of the baryon mass in and
        around galaxies, especially massive ones (e.g,, Crain et al. 2010). The X-ray wavelength range also contains major atomic transitions (e.g., all K-shell transitions of carbon through iron), providing key thermal, chemical, and kinematic diagnostics of the interstellar medium (ISM) in all phases (cold, warm, and
hot) and all forms (atomic, molecular, and dust).
X-rays are also energetic and penetrating, which may heat and evaporate dust grains/molecules and even companion stars in X-ray binaries, significantly affecting stellar and interstellar radiation. X-rays can often be observed through dense 
clouds with column densities up to $\gtrsim 10^{24} {\rm~cm^{-2}}$. Therefore,
X-ray observations are a powerful tool in examining the stellar and interstellar 
properties of galaxies. 

Substantial progress in understanding the X-ray emission from nearby galaxies
has been made over the past decade or so. \Chandra\ and \xmm\ X-ray 
Observatories, in particular, have provided a combination of  superb spatial and spectral resolutions (reaching $\sim 1^{\prime\prime}$ and $500 {\rm~km~s^{-1}}$ FWHM) as well as large collecting areas (up to $\sim 3000$ cm$^2$) over a broad energy range of 0.3-10 keV. These 
capabilities have enabled extensive studies of various
classes of X-ray sources: their luminosity functions, spectral characteristics, 
and relationships to other galactic properties such as the star formation 
rate (SFR) and total stellar mass. These are the topics that I am going to review, 
focusing on recent results.


\section{Discrete X-ray Sources}

{\underline{\it Active Galactic Nuclei (AGN)}}. 
AGNs are powered by the accretion of matter by supermassive black holes (SMBHs) 
from their surroundings. This process can be very efficient, energetically, converting up to about 10\% of the accreted mass into radiation and mechanical energy. For example, the energy
released from the growth of a SMBH can be about 100 times the gravitational binding energy of its host galaxy. This energy feedback, if well coupled to the surrounding medium, should then have enormous impact on galaxy formation and evolution. 

While much of the SMBH growth occurs at redshifts greater than 1, present AGNs typically have only moderate radiation luminosities $L_x \sim 10^{42}-10^{45} {\rm~ergs~s^{-1}}$ and appear in only about 10\% of nearby galaxies. Such AGNs are most easily identified in X-ray, since their luminosities are still considerably higher than those of other galactic sources. AGNs typically have intrinsic X-ray spectra that can be characterized by a power law with a photon index of $\sim 1.7$. However, a significant fraction of AGNs are known to be severely obscured or even Compton-thick to X-rays. Such AGNs may still be detected in reflected or reprocessed light, mostly in optical, mid-IR, and/or hard X-ray (from photon-ionized gas, hot dust, and/or  Fe florescence;  e.g., LaMassa et al. 2009).
Observations of the AGNs can also provide us with a better view of 
their feedback effects  on the immediate
surroundings. Their X-ray spectra typically
show a mixture of collisionally excited hot thermal plasma, characterized by numerous strong  resonance lines, and photo-ionized plasma, represented by strong forbidden 
lines and recombination continuum (e.g., Guainazzi \& Bianchi 2007). 
Interestingly, the X-ray luminosities required to model
such spectra sometimes far exceed those of the present
AGNs, indicating that they may have been substantially brighter 
in the recent past  (e.g., Wang et al. 2010) or that some other processes 
such as charge exchange (CX) between hot ions and neutral atoms may 
need to be considered (see later discussions; Liu et al. 2011b).

When a SMBH is in a low-$L_x$ ($\lesssim 10^{42}  {\rm~ergs~s^{-1}}$) or even 
moderate-$L_x$ state, the accretion is likely in the so-called radiatively 
inefficient (radio) mode (e.g., Yuan 2007). Much of
the gravitational energy of the accreted matter may be released in form of
mechanical energy (e.g., an accretion 
disk wind and/or jet; Omma et al. 2004). Such energy
release may play an important role in regulating the nuclear environment
of galaxies. A SMBH may also undergo intermittent
accretion episodes, resulting in cycles of heating and cooling of surrounding
gas (e.g., Yuan et al. 2009; Pellegrini et al. 2011).
Various studies are ongoing to explore the interplay between 
the SMBH accretion and the nuclear environment of very nearby galaxies 
($D \lesssim 10$ Mpc; e.g., Li et al. 2009, 2011).
For more distant galaxies, confusion with stellar sources typically 
becomes too severe to even identify individual low-$L_x$ AGNs with 
certainty, especially in galaxies with active nuclear SF.

{\underline{\it Sources Related to Recent Star Formation}}. 
A nearby galaxy, excluding its possible AGN, typically has 
$L_x \lesssim 10^{42}  {\rm~ergs~s^{-1}}$,
depending primarily on its mass and SFR.  In an active SF
galaxy, the luminosity is normally dominated by high-mass X-ray binaries (HMXBs),  which contain
neutron stars or stellar mass black holes accreting typically from stellar winds of companions
more massive than $\sim 5 M_\odot$. The spectrum of such a HMXB can be characterized  by a
power law with a photon index of  $\sim$ 1.2, a cutoff at $\sim 20$ keV, plus an Fe-K emission feature centered at 6.4-6.7 keV and with an equivalent width of 0.2-0.6 keV (White et al. 1983). The X-ray luminosity function (XLF) of HMXBs can be characterized by a power law with a differential slope of 1.5-1.6 over a very broad luminosity range of $10^{36} - 10^{40}  {\rm~ergs~s^{-1}}$ (Grimm et al. 2003; Persic \& Rephaeli 2007 and references therein). The XLF shows an exponential cutoff at $\sim 2 \times 10^{40} {\rm~ergs~s^{-1}}$ (Swartz et al. 2011) and seems
to flatten out again at $\gtrsim 10^{41}  {\rm~ergs~s^{-1}}$.

Much attention has been placed on so-called ultra-luminous X-ray sources (ULXs) ---
non-nuclear, point-like, X-ray sources with apparent isotropic  
$L_X \gtrsim 10^{39} {\rm~ergs~s^{-1}}$, which is roughly the Eddington luminosity 
of a stellar mass BH  (e.g., Swartz et al. 2011 and references therein). ULXs tend to have steep spectra with a mean power law photon 
index of $\sim 1.7$. There is on average $\sim 1$ ULX per 0.5 M$_\odot~{\rm~yr^{-1}}$ SFR in local galaxies. ULXs are clearly a heterogeneous population of X-ray sources, including very young supernova remnants (SNRs, due to the interaction of SN ejecta with dense circumstellar materials), as well as probably mildly beamed and/or super-Eddington HMXBs. Some of these sources, especially those rare ones with $L_X \gtrsim 10^{41}  {\rm~ergs~s^{-1}}$, may represent the so-called intermediate-mass BHs (IMBHs with $10^2 M_\odot \lesssim M_{BH} \lesssim 10^4 M_\odot$) undergoing sub-Eddington accretion. IMBHs may be formed from the collapse of Pop III stars and at centers of globular clusters (GCs) and dwarf elliptical galaxies, which may since being destroyed dynamically.

There are other young stellar objects and their remnants such as YSOs and colliding wind binaries, as well as normal SNRs. Such sources are typically quite faint 
individually ($\lesssim 10^{35}{\rm~ergs~s^{-1}}$); but collectively they can still 
account for a significant fraction of the X-ray emission from
galaxies, especially when the contribution 
from brighter sources has been excised: 2-10 keV
$L_x \approx 10^{38.2} {\rm~ergs~s^{-1}}[M_\odot{\rm~yr^{-1}]^{-1}}$ 
(Bogd\'an \& Gilfanov 2011). 

The {\sl total} X-ray luminosity of
young stellar objects and their remnants in a galaxy
is strongly correlated with the SFR: $L_x =10^{39.4} {\rm~ergs~s^{-1}}[M_\odot~{\rm~yr^{-1}]^{-1}}$ with a scatter of 0.4 dex (e.g., Mineo et al. 2011). 
With this correlation, accounting for its possible cosmological evolution 
(Dijkstra et al. 2011) as well as the scatter, one can estimate 
the SFR of a distant galaxy 
from its X-ray luminosity. Or if we know the SFR from its other proxies, 
we can then compare the predicted and observed luminosities to constrain a
potential AGN contribution.


{\underline{\it Sources from Old Stars}}. 
A galaxy with little recent SF can still contain copious X-ray sources, both discrete and diffuse. In this case, the brightest off-nuclear discrete sources are low-mass X-ray binaries (LMXBs), which contain neutron stars or stellar-mass BHs accreting from their
typically post-main-sequence, Roche-lobe-overflowing companions with 
masses  $\lesssim 1 M_\odot$. 
Therefore, LMXBs trace stellar populations older than $\sim 1$ Gyr. 

The average X-ray spectrum of LMXBs can be characterized by a power law with a photon index of $\sim 1.6$, considerably steeper than that of HMXBs. LMXBs at the high end of
their 0.3-10 keV luminosity range, $\sim 10^{39}  {\rm~ergs~s^{-1}}$,
tend to have even softer spectra and are similar to Galactic black hole X-ray 
binary candidates in their very high state (Irwin et al. 2003).

The XLF of LMXBs has a convex shape. It can be represented approximately by a power law with a differential slope of $\sim$ 2.0 in the luminosity range $L_X \sim 5 \times 10^{37}-5 \times 10^{38} {\rm~ergs~s^{-1}}$ (Gilfanov 2004; Kim \& Fabbiano 2004). The XLF steepens at higher luminosities, which appears
to be more significant with the 
increasing galaxy age (Kim \&  Fabbiano 2010); the XLF of young elliptical galaxies (a few Gyr old) is intermediate between that of typical old elliptical galaxies and that of star-forming galaxies. The overall steepening of the XLF with the luminosity can be explained by the mass transfer in binary systems with 
giants, which significantly shortens their stellar lifetime 
(Revnivtsev et al. 2011). In contrast, the majority of binary systems at luminosities $\lesssim 5 \times 10^{37}$ have main-sequence secondary companions (except for those with white dwarf donors), apparently responsible for the 
flattening of the XLF to a slope $\sim 1$ (Revnivtsev et al. 2011). 
The normalization of the XLF is 
proportional to the stellar mass, which can be estimated from the K-band luminosity 
using a color-based correction for the mass-to-light ratio of a galaxy
(Bell \& de Jong 2001). 
On average, the combined luminosity of LMXBs with individual $L_X \gtrsim 10^{37} {\rm~ergs~s^{-1}}$,
which can be detected in a nearby galaxy with a reasonable exposure of 
\chandra\ or \xmm,  is 
($8\pm0.5) \times 10^{39} {\rm~ergs~s^{-1}}$ per $10^{11} M_\odot$  (Gilfanov 2004);
the residual contribution down to $L_x \sim 10^{35} {\rm~ergs~s^{-1}}$
(typically $\lesssim 20\%$, especially in the $\lesssim 2 $ keV band) 
can be well estimated with the flat XLF.

The average properties of fainter X-ray sources, mostly cataclysmic variables (CVs) and active
binaries (ABs), which are typically in the range of $10^{30}<L_x<10^{34} {\rm~ergs~s^{-1}}$ (2-10 keV), have also been measured. Sazonov et al. (2006) show that the specific luminosity of such sources (per unit stellar mass) in the solar neighborhood
is $(2.0\pm 0.8)\times 10^{27}$ and $(1.1\pm 0.3)\times 10^{27} {\rm~erg ~s^{-1}~ M_\odot^{-1}}$. The inferred total contribution of ABs and CVs to 
the 2-10 keV luminosity of the Milky Way is 
$\sim 2 \times 10^{38}  {\rm~ergs~s^{-1}}$, or ~3\% of the integral luminosity of LMXBs. The specific luminosity has also been estimated for nearby low-mass
early-type galaxies (M32 and NGC 3379; Revnivtsev et al. 2008) and is shown to  be a factor 
of $\sim 2$ lower than the solar neighborhood value. The exact cause of
this difference is yet to be identified. One possibility is that it represents
the genuine difference in the binarity of stars, resulting from different
stellar formation and evolution environments.

\begin{figure}[b]
\vskip -0.4cm
\begin{center}
  \centerline{
  \includegraphics[width=5.5in]{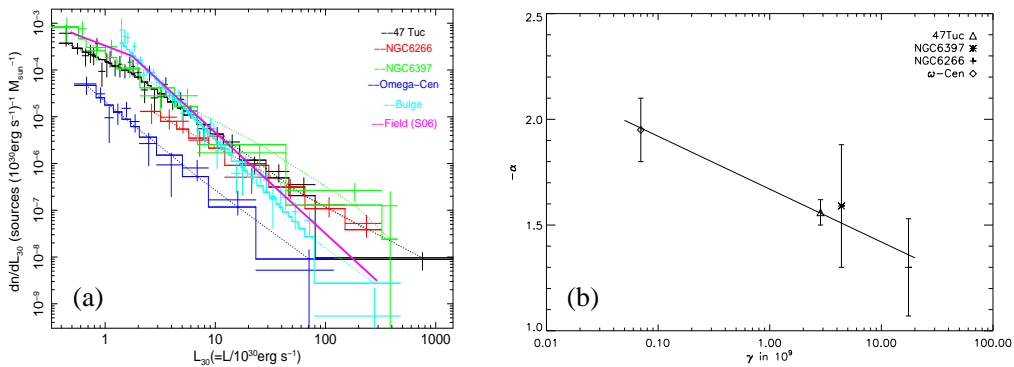} 
}
\vskip -0.2cm
\caption{(a) Specific XLFs of four GCs and a Galactic bulge field, as well as the best-fit
broken power law model for sources in the solar neighborhood (Field; 
Xu, Wang, \& Li 2011). The incompleteness and bias of the source detection have been
approximately corrected, while the contributions from interlopers have been subtracted
statistically from the data. (b) Correlation between the power law slopes
of the LFs and the mass-averaged encounter rate ($\gamma$) of the GCs, demonstrating the
importance of the dynamical effect on formation and evolution of binaries.
 }
   \label{f1}
\end{center}
\end{figure}


Indeed, the environment-dependence of the XLF is clearly present 
for LMXBs (e.g., Zhang et al.
2011 and references therein). The XLF shape of sources in globular clusters 
(GCs) is shown to be significantly different from that in the field,
especially in $10^{35} - 10^{37} {\rm~ergs~s^{-1}}$.
Over this luminosity range (relative to more 
luminous ones), the fraction of sources in GCs is about 4 times 
less than in the field. 
A similar result is also obtained in a recent study of 
the XLFs of sources over a substantially lower luminosity range (e.g., 
Fig.~\ref{f1}; Xu, Wang, \& Li 2011). The relatively high luminosity 
range ($L_X \gtrsim 10^{31}$) should be
dominated by CVs. The specific population of such sources 
in a GC depends on its dynamical state and can be greater than that 
in the Galactic bulge or in the solar neighborhood. Interestingly, 
the source population, particularly in the lower luminosity range, 
is generally smaller 
in the GCs (especially in those dynamically less evolved ones) 
than in the bulge and field. To explain this result, 
one probably needs to invoke both the dynamical
formation of binaries and the assumption that the initial binary 
fraction of stars is very small in GCs, consistent with existing 
optical observations (e.g., Davis et al. 2008). 
The dynamical effect has also been shown to be important
in explaining the increasing specific LMXB population in the central bulge
of M31 (Zhang et al. 2011). 
In regions outside of GCs and Galactic nuclear regions,
however, we can now reasonably determine and subtract the faint stellar 
X-ray contribution (Revnivtsev et al. 2008; Zhang et al. 2011) to facilitate 
the measurements of truly diffuse X-ray emission from galaxies.

\section{Diffuse Soft X-ray Emission}

Diffuse soft X-ray emission has commonly been used to trace various types of galactic
feedback in nearby starburst and normal galaxies, as well as the cooling of 
hot gaseous halos or coronae resulting from the galaxy formation (accretion) 
processes  (e.g, Crain et al. 2010).  Assuming an origin of this emission in
optically-thin thermal (collisionally-excited) hot plasma, one may estimate its 
mass, energy, and chemical contents and even their outflow or accretion 
rates from a galaxy. 

\begin{figure*}[!bth]
\vskip -0.4cm
  \centerline{
  \includegraphics[width=5.2in]{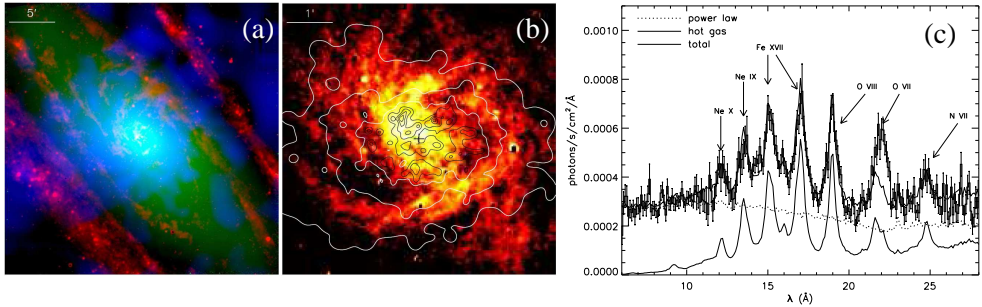} 
}
\vskip -0.2cm
  \caption{\footnotesize  {\bf (a)} Tri-color image of the spheroid region of M31: 
{\sl Spitzer}/MIPS 24 $\mu$m emission ({\sl red}),
   2MASS K-band emission ({\sl green}), and
   {\sl Chandra} 0.5-2 keV emission of truly diffuse hot gas ({\sl blue}; 
Li \& Wang 2007).
    {\bf (b)} Intensity contours of the  diffuse emission overlaid
    on an H$_{\alpha}$ image in the M31 central region. A detailed 
analysis of the data shows evidence for strong interaction between the hot 
and cool gas phases, which may provide a mechanism for starving the nucleus
as marked by the plus sign (Li et al. 2009). 
{\bf (c)} An \xmm\ Reflection Grating Spectrometer (RGS) spectrum of the M31 central region (Liu et al. 2010). The curves 
show the fit with a single-temperature thermal plasma, plus a power-law that characterizes 
the discrete source contribution. Note the intensity excess of the \ovii\ K$\alpha$ triplet
above the model, giving a strong indication for emission from the CX between
hot ions and neutral atoms. }
\label{f2}
\end{figure*}

{\underline{\it Emission from Galactic Spheroids}}.
While galactic coronae associated with giant elliptical galaxies have been studied
extensively
(e.g, Sun et al. 2007; Mulchaey \& Jeltema 2010; Pellegrini 2011), 
the very detection of truly diffuse X-ray 
emission in and around low-$L_x$ stellar spheroids (galactic bulges or low- to 
intermediate-mass ellipticals) becomes possible only recently, thanks largely to
the calibration of the stellar $L_x$ to mass ratio, as discussed above. A good example 
of such a detection is the bulge of M31 (Fig~\ref{f2}a; 
Li et al. 2007;  Bogd\'an \& Gilfanov 2011). The large-scale X-ray 
emission shows a bipolar morphology along the minor axis of the galaxy, 
indicating an outflow of hot plasma, which is most likely driven by Ia SNe because there is essentially no AGN and no
recent SF in the bulge. But the X-ray luminosity accounts for only $\sim 2\%$ of the Ia SN 
mechanical energy input in the bulge. There is no bulge-wide distributed
cool gas to potentially consume or convert the energy into other forms. Therefore,
the outflow is actually required to explain the lack of the accumulation of the stellar feedback, not only the energy, but the mass (mainly from redgiant winds and planetary nebulae) and the iron (from Ia SNe, about 0.7 $M_\odot$ each)
as well. Theoretically, it has been shown that such outflows play an important role in maintaining large-scale hot gaseous halos around galactic spheroids (Tang et al. 2009).

The dynamical state of a corona around a galactic spheroid depends on its gravitational mass and environment (e.g., Wang et al. 2004; Owen et al. 2006; Sun et al. 2007, 2010; 
Mulchaey \& Jeltema 2010; We\'zgowiec et al.  2011; Tang et al. 2009; Lu \& Wang 2011). 
The corona of an intermediate-mass spheroid fast moving in a rich cluster, for example, most likely represents an Ia SN-driven outflow semi-confined by the ram pressure of the intracluster medium (ICM; Lu \& Wang 2011). The ram pressure affects the size and lopsidedness of the coronae. Because of the semi-confinement, the outflow should also be subsonic. The hot gas density increases with the ICM thermal pressure,
which may lead to the compression of cool gas clouds, if present, and hence the formation of stars (Young \& Scoville 1991; Bekki \& Couch 2003). Such density increase also enhances radiative cooling of the hot gas, which may fuel central supermassive black holes, explaining why the frequency of active galactic nuclei observed in clusters tends to be higher than that in the field (e.g., Hart, Stocke \& Hallman 2009) and is apparently responsible for a substantially higher surface brightness of the X-ray emission detected from corona in the cluster environment. Furthermore, the total X-ray luminosity of a corona depends on the relative importance of the surrounding thermal and ram pressures. These environment dependencies should at least partly explain the large dispersion in the observed diffuse X-ray luminosities of spheroids with similar stellar properties. Furthermore, an outflow powered by the distributed feedback can naturally produce a positive radial gradient in the hot gas entropy, mimicking a cooling flow. 

However, the soft X-ray emission in the inner regions of  spheroids is still far 
from being well understood.
For example, the low temperature and metal abundance of hot plasma, as inferred from
the existing spectral studies, are still difficult to be reconciled with models and simulations (e.g.,
Tang \& Wang 2010). Even the very assumption that the X-ray line emission originates 
predominantly in optically-thin thermal plasma is problematic (Liu et al. 2010).
This becomes apparent
when the so-called G [$=(f +i)/r$, forbidden+inter-combination to resonance line]
ratio of the K$\alpha$ triplet of such a diagnostic He-like ion as \ovii\ is analyzed.
For an optically-thin thermal plasma with a temperature of a few $10^6$ K
and in collisional ionization equilibrium, G should be considerably less than one
(e.g., Porquet et al. 2010). Fig.~\ref{f2}c presents an RGS spectrum of the M31 
central region (Fig~\ref{f2}b), showing a clear intensity excess above a simple 
thermal plasma model 
fit at $\sim 22$ \AA. This excess is due to the presence of the 
strong forbidden line 
(at 22.1 \AA), relative to the resonant one (21.6 \AA) of the 
\ovii\ K$\alpha$ triplet. Furthermore,
the excess is shown to extend throughout much of the cross-dispersion range
of the RGS observations and seems to be correlated well with the distribution of the cool gas
(Fig.~\ref{f2}b). One natural explanation is that the excess or the strong forbidden
line represents
the contribution from the CX between the two gas phases. 

{\underline{\it Emission from Active SF Galaxies}}.
Similar conclusions are also reached in studies of diffuse soft X-ray 
emission from active 
SF galaxies. Such emission has been mapped out for many face-on galaxies
(e.g., Tyler et al. 2003; Doane et al. 2004;  Owen \& Warwick 2009; 
Kuntz \& Snowden 2010). The emission is 
closely correlated with recent SF. 
The study of M101, for example, shows that the bulk ($\gtrsim 80\%$)
of the emission is associated with SF regions with ages $\lesssim 20$
Myr, but is not due to individual SNRs,
as identified in optical and radio (e.g., Kuntz \& Snowden 2010). 
The emission traces diffuse hot plasma, 
which can naturally be heated by fast stellar winds and SNe 
of massive stars. This is most
vividly demonstrated by the emission from within cool gas cavities
around massive stellar clusters, as in the 30 Doradus nebula
(Wang et al. 1999). 
The X-ray CCD spectra of the emission from galaxies can typically be approximately
fitted with an optically-thin thermal plasma model with
two characteristic temperatures of  $\sim 0.2\pm0.1$ and $0.7\pm0.1$ keV;
the low-temperature component typically accounts for the bulk of the observed flux. 
However, the total luminosity of the emission accounts for a very 
small fraction ($\lesssim 1\%$) of the expected mechanical energy input rate from
fast stellar winds and SNe of massive stars. The estimated
thermal energy of the plasma is also substantially smaller than the expected  
total energy input. Although a fraction of the energy may be carried
away by cosmic rays or converted to radiation in other wavelength
bands, outflows from recent SF regions (i.e., blown-out
superbubbles or galactic chimneys) are expected, venting chemically-enriched hot plasma into the
large-scale halos of galaxies. Indeed, extraplanar soft X-ray emission is observed around
active SF galaxies (Wang et al. 2001, 2003; T\"ullmann  2006; Wang 2010 
and references therein; Li \& Wang 2011; Anderson \& Bregman 2011). Such emission also often appears to be 
correlated with extraplanar H$\alpha$-emitting and/or dusty
features. This correlation has led  to the conclusion that the bulk of the emission arises from the interaction between the outflowing hot plasma and entrained or pre-existing cool gas clouds, rather than from the superwinds themselves (e.g., Strickland \& Heckman 2009). 

To determine the nature of the interaction, we have recently examined the 
spectroscopic properties of the soft X-ray emission from nine
nearby active SF galaxies, which show no significant AGN activities 
(Liu et al. 2011a,b), using high spectral resolution RGS
data with good signal-to-noise ratios. 
The RGS spectra of these galaxies typically show large G ratios of 
the OVII K$\alpha$ triplet, indicating that a substantial
fraction of the emission cannot simply arise from collisionally-excited hot 
plasma, as commonly assumed. The bulk of the OVII K$\alpha$
line emission is consistent with an origin in the CX between highly
ionized ions (i.e., OVIII) and 
neutral atoms (e.g., HI and HeI); other possible mechanisms for
producing a high G ratio, such as a collisional non-equilibrium-ionization 
recombining/ionizing plasma, are not favored, although they cannot be 
completely ruled out. In the central region
of M82, for example, the putative CX contribution  to the
key He-like K$\alpha$ triplets is on average $\gtrsim 50\%$, decreasing with the increasing
ionization states (i.e., the G ratio is smaller for NeIX and Mg XI than for OVII; Liu et al.
2011a). To quantify the true contribution of the CX, one needs to have
a spectral model of it, which is not yet available. Nevertheless, 
the joint analysis of multiple emission lines indicates that the thermal
plasma is substantially hotter, probably
consistent with the high-temperature component indicated in
the 2-temperature characterization mentioned above.  The decomposition of 
the CX and thermal contributions will also be essential to a correct estimation of the 
metal abundances of the plasma (Liu et al. 2011a). 

The low X-ray emission efficiency of the galactic feedback  ($\sim
1\%$) has strong implications for understanding the formation and evolution
of galaxies.
The efficiency shows no significant correlation with the stellar mass
or gas content of a galaxy (Li \& Wang 2011). There is also little evidence that the
feedback energy is consumed at other wavelengths. For example, Grimes
et al. (2009) show that O VI emission is not generally detected in
their sample of starburst galaxies, suggesting that radiative cooling
or turbulent mixing is not significant in draining energy from
galactic winds. One may thus conclude that the bulk of the energy is
gone with the winds and is released to the halos of galaxies,
a topic that certainly deserves further attention.

\section{Summary and Conclusions}
I have reviewed some basic properties of various types of X-ray sources, particularly 
in relation to the spectral energy distribution (SED)
of galaxies. The key points are as follows:
\begin{itemize}
\item AGNs are typically the brightest X-ray sources in galaxies and can be detected relatively easily
even at high redshifts with little confusion. Such detection is important in 
modeling the SED of galaxies. Even low-$L_x$ AGNs, as typically
seen in nearby galaxies, can be important in regulating galactic 
nuclear environments.

\item In nearby galaxies, the bulk of HMXBs and LMXBs can readily be detected. Their
overall galactic populations are well correlated with the SFR and stellar mass and 
can also be independently estimated (e.g., with infrared observations).
The comparisons of these independent estimates can then 
be used to constrain the 
presence of AGNs in distant galaxies, even if they are not well resolved. 

\item The X-ray source population  of a stellar system is particularly sensitive to 
its binary fraction. Thus X-ray observations provide a powerful tool
 to probe the initial binary fraction and dynamical evolution 
of stellar clusters and even galactic spheroids.

\item Diffuse X-ray emission traces the galactic feedback, primarily due to massive SF (via fast stellar winds and core-collapsed SNe) 
in galactic disks, to old stars (mass-loss
and Ia SNe), and to possible AGNs. However, the  luminosity of the 
X-ray emission from a galaxy typically accounts for $\lesssim 1\%$ of the feedback
energy. The bulk of the energy is released in 
outflows of hot plasma and cosmic-rays/magnetic fields 
from stellar spheroids, as well as from SF regions. 

\item A substantial fraction of the diffuse X-ray emission may arise from the CX 
at the interface between hot plasma and cool gas. The CX, 
in principle, can be used as a new tool to probe the thermal, chemical, 
and kinematic properties of hot plasma tracing  
various high-energy feedback processes in the galaxies.
\end{itemize}

While much can still be done with the existing X-ray observatories, we will
also have new tools in the near future. {\sl NuSTAR} (scheduled to be launched
in late 2012) will offer the first imaging capability 
in the 8-80 keV range, while {\sl Astro-H} (2013) will
provide the calorimeter-based imaging spectroscopic capability in 
the 0.5-10 keV range. Furthermore, {\sl eROSITA} (2013) will conduct
the first all-sky X-ray CCD imaging survey over the 0.3-10 keV range, 
while {\sl GEMS} (2014) will enable 
polarization measurements in the 2-10 keV range. With these new capabilities, we can expect
a substantial advance in our understanding of X-ray sources in galaxies.

I am grateful to the organizers of the IAU Symposium for inviting me to give
a review talk on this topic, which the present article is based on, and to 
my students and collaborators for their contributions to some of 
the work described above. My research is largely supported by NASA via various 
grants.

\end{document}